\documentstyle [12pt,epsf]{article}

\begin{document}

\begin{titlepage}

\null
\begin{flushright}
CERN-TH.7286/94
\end{flushright}
\vspace{20mm}

\begin{center}
\bf\Large{Simulations of dynamically triangulated gravity \\
       -- an algorithm for arbitrary dimension}
\end{center}

\vspace{5mm}

\begin{center} 
\bf{S.Catterall\footnote{Permanent address: Physics Department, Syracuse
University, Syracuse, NY 13244}}\\
        TH-Division, CERN CH-1211,\\
        Geneva 23, Switzerland.
\end{center}

\vspace{10mm}

\begin{abstract}
Recent models for discrete euclidean quantum gravity
incorporate a sum over simplicial triangulations. We describe an 
algorithm for simulating such models in
arbitrary dimension. As illustration we show results from
simulations in four dimensions. 
\end{abstract}

\vfill
\begin{flushleft}
CERN-TH.7286/94\\
June 1994
\end{flushleft}

\end{titlepage}

\section*{Introduction}

In recent years there has been considerable interest in studying
statistical systems whose partition functions encorporate a sum
over simplicial triangulations. Initially, efforts focused on two
dimensional models which were proposed as discrete  
regularisations for string theory out of the critical dimension 
\cite{mig,amb,dav}. Plausibility arguments were given to suggest
that the sum over lattices, in some scaling limit, would generate
the effects of a nonperturbative inclusion of fluctuations in the
worldsheet geometry. Later, a framework was developed \cite{kpz,ddk}
which, in certain simple cases, allowed many of the important features of
a system coupled to two dimensional gravity to be derived from
a knowledge of the theory in flat space. 

The results of these continuum analyses were in complete agreement
with calculations and numerical simulations of the triangulated
models \cite{mig2d,jurkis,usis,clivepot,kaz,puregrav} 
and lent strong support to the lattice prescription. The continuum
methods used to analyse the simple models (including pure two
dimensional gravity) appear to break down for strings in
physical dimensions. This has motivated a variety of numerical
studies of the discrete models (which are well defined everywhere)
with some interesting results \cite{multis,multipotts,multiamb,xy}.

In addition, the basic idea of summing over simplicial triangulations
to generate a path integral for quantum gravity has been extended
to three \cite{3dgross,3damb,3dmig,3dboul} and four dimensions 
\cite{4damb,4dsteen,4dmig,4dbrug,4dus}.
Whilst these initial simulations are
rather exploratory, the results for four dimensions are particularly
exciting, as they seem to hint at a nonperturbative fixed point 
in the quantum
theory.  It is possible that the problems associated with the
nonrenormalisability of Einstein gravity might be evaded in any
continuum theory constructed in the vicinity of this new fixed point. 

In this paper we present an algorithm for Monte Carlo simulation
of these dynamically triangulated models, which is constructed in
such a way as to make trivial the dependence of the
code on the manifold dimension $d$.

\section*{Model}

We will be considering the problem of estimating a partition
function of the form
\begin{equation}
Z\left(\kappa_0,\kappa_d\right)=\sum_{T\left(S^d\right)}
e^{-S\left(\kappa_0,\kappa_d,T\right)}
\end{equation}

The summation goes over all simplicial triangulations $T$ of the
sphere in d dimensions $S^d$. A simplicial triangulation is
specified by a set of $d$-simplices (sets of $d+1$ labelled points)
which are associated uniquely in pairs via their $d-1$-dimensional
faces, in such a way that the neighbourhood of any point is
homeomorphic to a $d$-dimensional ball. In two
dimensions $d=2$, the fundamental building blocks are
$2$-simplices (triangles) which are glued together along
their $1$-dimensional faces (links) in such a way that
two points (vertices) are connected by at most one link and the
end points of all links are distinct. Similarly, a three
dimensional simplicial manifold is built out of $3$-simplices
(tetrahedra) such that a given $2$-dimensional face (triangle)
is associated with exactly two tetrahedra. The restriction
to manifolds ensures that every subsimplex is nondegenerate and
unique. Analogously, in
four dimensions, the triangulation consists of hypertetrahedra associated
in pairs via their tetrahedral faces. Again, the manifold condition effectively
eliminates any degeneracies in the subsimplices.

The action $S$ for
dimensions $d\le 4$ can be taken to depend on only two
coupling constants $\kappa_0$, $\kappa_d$ related to the
bare Newton and cosmological constants respectively. They are
conjugate to the total number of $d$-simplices $N_d$ and points $N_0$
($0$-simplices) 
for a given triangulation $T$.

\begin{equation}
S\left(\kappa_0,\kappa_d\right)=\kappa_d N_d - \kappa_0 N_0
\end{equation}

The numerical
evaluation of this partition function $Z$
(and expectation values computed from it) is effected by a Monte Carlo
procedure which generates a random walk in the space
of all such triangulations by a sequence of
local `moves' or deformations. The ones commonly used correspond to
the replacement of an $i$-dimensional subsimplex (i.e a subset of $i+1$
points within a simplex) by its `dual' $d-i$ subsimplex (see, for example,
\cite{gross}). In order
that this move preserve the manifold structure of the triangulation, there
will be an associated change in the number and identity of
neighbouring simplices and subsimplices. These moves have been shown to be
ergodic (at least when $d\le 4$) in \cite{gross}. The latter statement
implies that, at least in principle, a set of such moves can
transform any such triangulation into any other of the same Euler
character $\chi$. The topological invariant $\chi$ is defined for
triangulations as

\begin{equation}
\chi=\sum_{i=0}^d \left(-1\right)^iN_i
\end{equation}

However, in the case of $d=4$, it appears that the typical number
of such moves may increase very rapidly with volume (number of
$d$-simplices). This may place important constraints on the `practical'
ergodicity of the numerical simulations \cite{ergod}. Indeed, there
is some recent numerical evidence that in this case the triangulation space may
grow factorially with volume \cite{bound}.

Clearly, there are $d+1$ possible moves which may be labeled by 
the dimension of the subsimplex $i$ central to the move. The definition
of the $i$-move requires that the order of a given $i$-subsimplex 
(the number of $d$-simplices associated with it) be exactly $d+1-i$.
Such a subsimplex will be referred to as a {\it legal} subsimplex.
Subsequent upon finding such a legal subsimplex it is necessary to
test whether substituting it by its `dual' will lead to a bona fide
triangulation which satisfies the manifold restriction. Such a move
is referred to as geometrically allowed. Finally, for such a
geometrically allowed move, the change in the action is computed and
the update subjected to a metropolis test. Together with
an explicit detailed balance condition, this procedure, under repeated
iteration, will guarantee that the configurations approach
a static distribution governed by the Boltzmann weight. The details
of this algorithm are given in the next section.

\section*{Algorithm}

The recipe for generating legal moves is as follows. Select a move type $i$
at random. Then choose a simplex and one of its $i$-subsimplices
($\left(i+1\right)$ vertex labels) also
at random. Using a local procedure find the order of this subsimplex
(i.e the number of simplices to which it is associated). If $O\left(i\right)\ne
\left(d+1-i\right)$ go back and select another move type. 

The details of the neighbour search are organised as
follows. Denote the labels of the $\left(d+1\right)$ points making
up a simplex containing the subsimplex $i$ in question by $a_0\ldots a_d$.
Examine all neighbour simplices which are associated with this
simplex by any face containing the $i$-subsimplex. There are $d-i$ of
these and each contains one vertex which is not in the original
simplex. If the move is to be `legal' (i.e the subsimplex $i$ is
of order $d+1-i$) then this extra vertex must be the same in all these
$d-i$ cases and can be denoted $a_{d+1}$. The slight exception
to this picture corresponds to barycentric node
insertion ($i=d$) where the extra vertex is a new label and
no searching is required. 

It is now
convenient to relabel the $d+2$ vertices central to the move 
in such a way that the $i+1$ points that define the
subsimplex are arranged from $a_0$ to $a_i$, the content of the
$i$-move may then be seen from the following construction
\begin{equation}
\overbrace{a_0\ldots a_i}a_{i+1}\ldots a_da_{d+1}\to a_0\ldots a_i
\overbrace{a_{i+1}\ldots a_da_{d+1}}
\end{equation}
The $d+1-i$ initial state simplices (the lefthand side of this equation) are
constructed by pairing the common subsimplex vertices (indicated by
the brace) with $d-i$ selected from the $d-i+1$ other vertices. The final
state is now gotten by identifying the
points $a_{i+1}\ldots a_da_{d+1}$ as the new common subsimplex
vertices (as indicated by the shift of the brace). The vertices needed
to make up the $i+1$ final state simplices are just $i$ selected from
the $i+1$ remaining $a_0\ldots a_i$.

In order that the new simplicial complex still corresponds to a
triangulation of a manifold, it is necessary to check that the potential
new simplices and subsimplices introduced by such a move are not
already present in the triangulation. In effect, this means that the
extra vertex $a_{d+1}$ must not already exist in any simplex
associated to the subset $a_{i+1}\ldots a_d$. To check for this
a local search is carried out on all simplices which contain this
subset. The nearby simplices are explored by moving out on
faces containing this subset, with simplices being flagged and
removed from the search list when they have been examined once.

Once this manifold condition has been checked, the update is
treated by the usual Metropolis test and the triangulation updated
if necessary. In order that the simulation produce the correct
Boltzmann probability density we have chosen to encorporate strict
detailed balance into the algorithm (see \cite{4dbrug}). 
Denoting the probability
of transition between one state or triangulation $\alpha$ to another $\beta$
via some subsimplex move $i$
by $\tau\left(i,\alpha ,\beta\right)$,  
detailed balance requires
\begin{equation}
P\left(\alpha\right)\tau\left(i,\alpha ,\beta\right)=P\left(\beta\right)
\tau\left(d-i,\beta ,\alpha\right)
\end{equation}
$P\left(\alpha\right)$ is the usual factor
\begin{equation}
P\left(\alpha\right)=e^{-S\left(\alpha\right)}
\end{equation}
In practice, the transition
probability factors into a product of 
probabilities to select the initial and final states 
-- $\eta\left(i,\alpha\right)\phi\left(d-i,\beta\right)$
, together with a 
piece $t\left(i,\alpha ,\beta\right)$
dependent on the change in action $S$. Here, a choice of initial
state produces a unique final state so $\phi=1$. 

In practice, an attempted update starts with a random selection of
the move type $i$ followed by a random selection
of a simplex. The probability then of selecting a given $i$-subsimplex
is then ${O\left(i\right)\over N_d\left(\alpha\right)}$ where $O\left(i
\right)$ is the order of the subsimplex in the triangulation $\alpha$. 
For a legal move $O\left(i\right)=\left(d+1-i\right)$. If
the $i+1$ vertices are then drawn at random from this simplex, the total
probability of selection is
\begin{equation}
\eta\left(i,\alpha\right)={1\over d+1}{1\over N_d\left(\alpha\right)}
{\left(d+1-i\right)\over {d+1\choose i+1}}
\end{equation}
It is elementary
to then see that the inverse move $\eta\left(d-i,\beta\right)$ differs
only by the number of simplices $N_d\left(\beta\right)$.
Thus eqn. 5 reads
\begin{equation}
e^{-S\left(\alpha\right)}{1\over N_d\left(\alpha\right)}
t\left(i,\alpha ,\beta\right)=
e^{-S\left(\beta\right)}{1\over N_d\left(\beta\right)}
t\left(d-i,\beta ,\alpha\right)
\end{equation}
This relation is
then satisfied by choosing the reduced transition matrix 
$t\left(i,\alpha ,\beta\right)$ 
to have the simple form 
\begin{equation}
t\left(i,\alpha,\beta\right)={1\over 1+\left(1+{\left(2i-d\right)\over N_d
\left(\alpha\right)}\right)
e^{S\left(\beta\right) -S\left(\alpha\right)}}
\end{equation}
The change in action only depends on the order $i$ of the subsimplex
move (since the total change in the number of simplices is $2i-d$). 
\begin{equation}
S\left(\beta\right) -S\left(\alpha\right) = 
\kappa_d\left(2i-d\right)-\kappa_0\left(\delta_{i,d}-
\delta_{i,0}\right)+\gamma\left(2i-d\right)\left(2\left(N_d-V\right)+2i-d\right)
\end{equation}
The supplementary term with coefficient $\gamma$ acts to 
control the volume fluctuations so
that with a tuning of $\kappa_d$ we can simulate a quasi microcanonical
ensemble of fixed volume $V$. Typically the coupling $\gamma$ is
taken small $\gamma=0.005$. We have verified that expectation values
computed at small $\gamma$ are independent of $\gamma$ but possess
statistical errors that grow as $\gamma\to 0$.

To simulate a lattice with volume $V$ we tune the bare cosmological coupling
$\kappa_d$ during equilibration according to a formula which 
follows from steepest descent evaluation of the partition function

\begin{equation}
\delta\kappa_d=2\gamma\left(\left\langle N_d\right\rangle-V\right)
\end{equation}

\section*{Data structures and practical considerations}

The code is written in C in order to handle dynamic memory allocation.
A structure of type SIMPLEX is defined which contains both
an array of labels for its vertices and an array of pointers to
the neighbour simplices. The pointer to a
neighbour simplex is stored with a local array index
identical to the vertex which does not appear in the
face separating the two simplices. In addition each simplex contains
a logical flag which may be set and unset during simplex searching
operations to prevent a given simplex being used more than once. Finally
the sum of its labels is also stored, as this allows for a fast
calculation of the opposing vertex of a new simplex neighbour across
a given face \cite{4dbrug}.
Sequences of functions allocate and delete simplices dynamically
and update the pointer fields of simplices neighbour to a move.
To handle the node insertion and deletion moves, a stack of
`used' vertex labels is maintained. If a node insertion is
attempted, the new label is drawn from the top of this stack, unless the
stack is empty in which case the total node number is
incremented. Conversely deleted nodes are placed on the stack.
The stack itself is managed as a linked list.

The total storage is of order $4\left(2d+12\right)V$ bytes for a $V$
simplex simulation. This equates to approximately $0.6$ Mbyte for an $8000$
simplex lattice in four dimensions. 

The update time for $V=8000$, $d=4$
at $\kappa_0=0.0$ is of order $4000$
microsecs per {\it accepted} elementary move on a HP-735
workstation. One Monte Carlo sweep is defined as $V$ attempted, legal
subsimplex moves. At $V=8000$ our sweep time is 1.4 secs -- this
equates to 180 microsecs per attempted move and hence our
average acceptance rate is approximately $5\%$. Sweep times decrease
monotonically 
with increasing $\kappa_0$ 
due to the decreased connectivity of the lattice yielding
significantly faster local search times. At $\kappa_0=2.4$ (close
to criticality for $V=8000$) the sweep time is just 0.9 secs with
a correspondingly smaller update time per accepted move.

The CPU time per attempted move increases with volume in $d=4$ and
for small $\kappa_0$: for
$V=4000$ it is 120 microsec, at $V=8000$ it is 180 microsec and
for $V=16000$ it has reached 230 microsec. However for sufficiently
large $\kappa_0$ it is essentially constant. 

\begin{figure}
\begin{center}
\leavevmode
\epsfxsize=400pt
\epsfbox{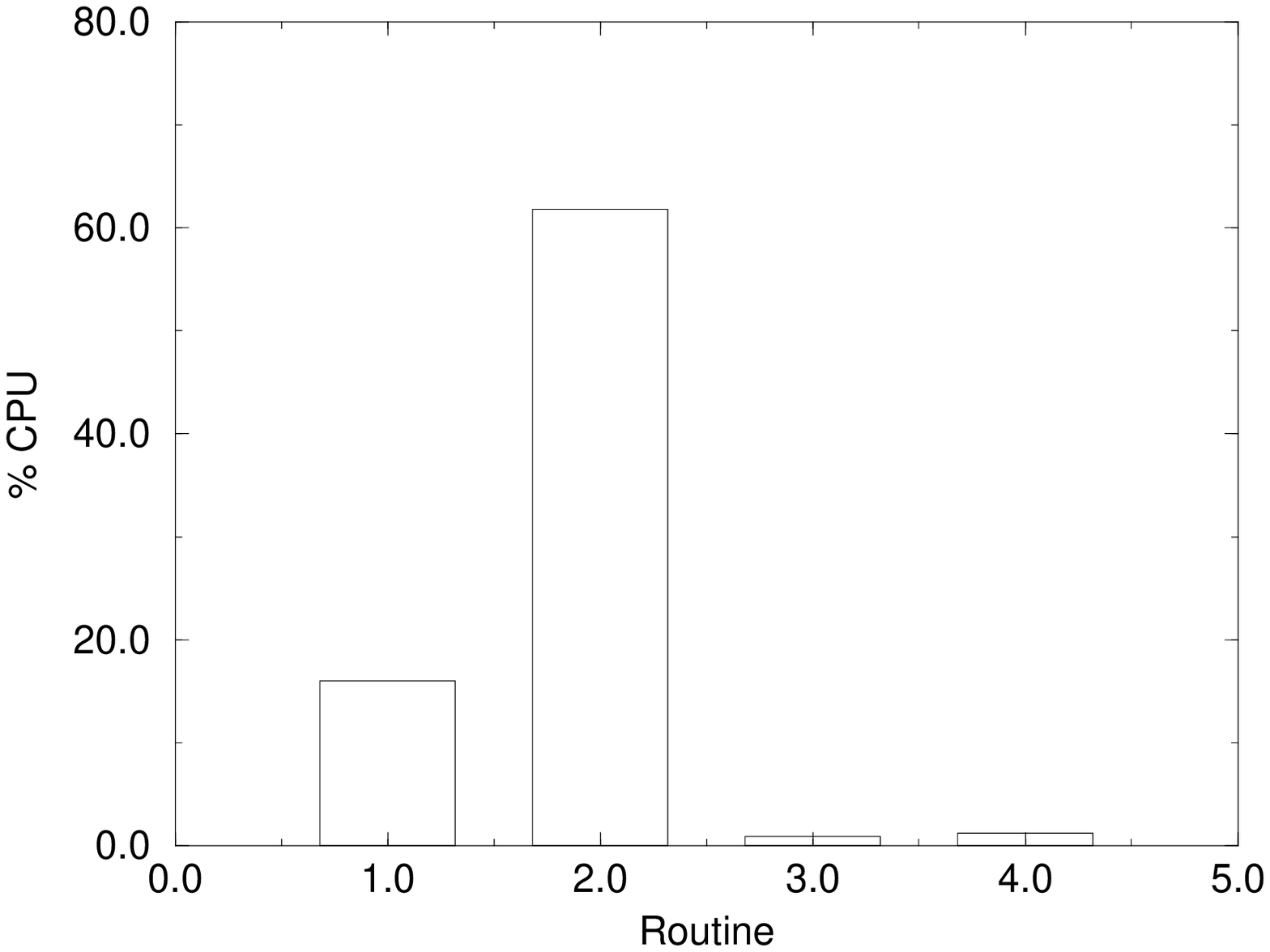}
\caption{Execution profile $d=4$,$V=8000$,$\kappa_0=0$}
\label{fig0}
\end{center}
\end{figure}

Fig. \ \ref{fig0} illustrates a typical execution profile for the code
with $d=4$ and $V=8000$, giving the percentage CPU time 
spent in the most important
routines which are labelled according to

\begin{enumerate}
\item Searching for legal subsimplices
\item Checking the geometric constraints
\item Computing the metropolis test
\item Updating the lattice structures
\end{enumerate}

Clearly, the program is dominated by the searching required to
check that a proposed move does not violate the geometric restriction
to manifolds.

\section*{Characteristic output}

\begin{figure}
\begin{center}
\leavevmode
\epsfxsize=400pt
\epsfbox{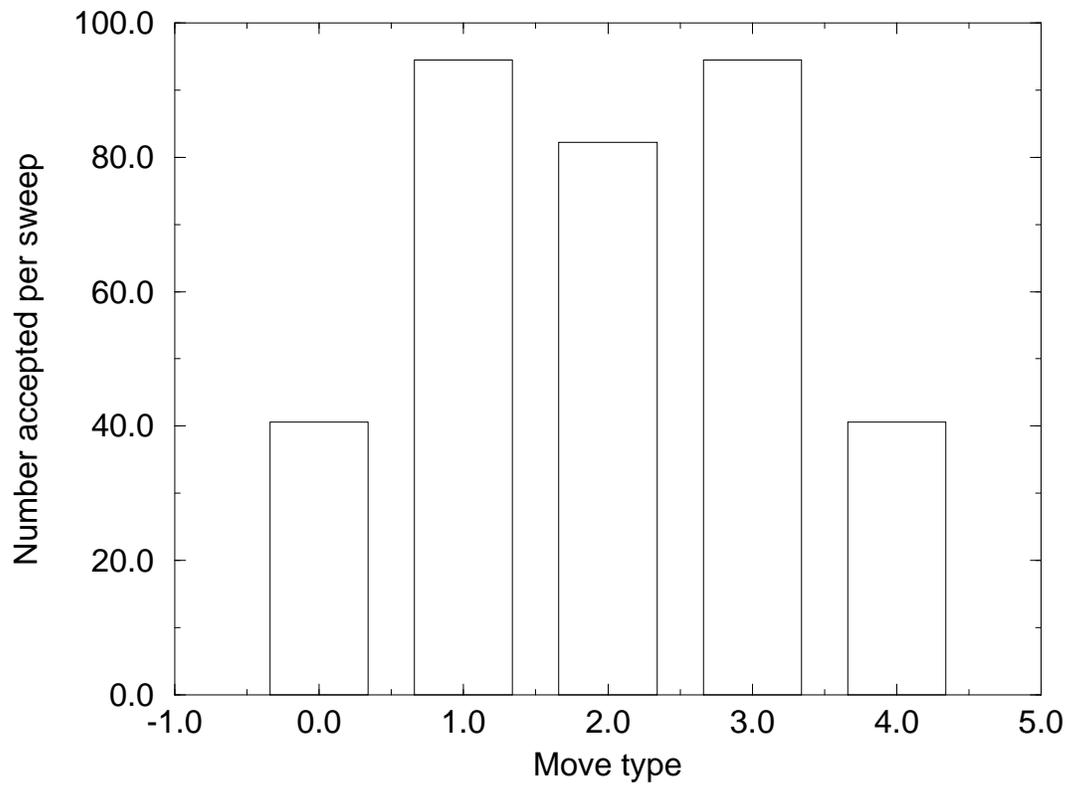}
\caption{Number accepted i-moves per sweep $d=4$, $V=8000$, $\kappa_0=0$}
\label{fig1}
\end{center}
\end{figure}

Fig. \ \ref{fig1} is a histogram illustrating $A\left(i\right)$ the number of
accepted moves of type $i$ per sweep for a four
dimensional lattice of volume $V=8000$
at zero node coupling. The manifest symmetry about $i=d/2$ is
a crude check of detailed balance -- there are as many moves of type
$i$ as inverse moves of type $d-i$. 

\begin{figure}
\begin{center}
\leavevmode
\epsfxsize=400pt
\epsfbox{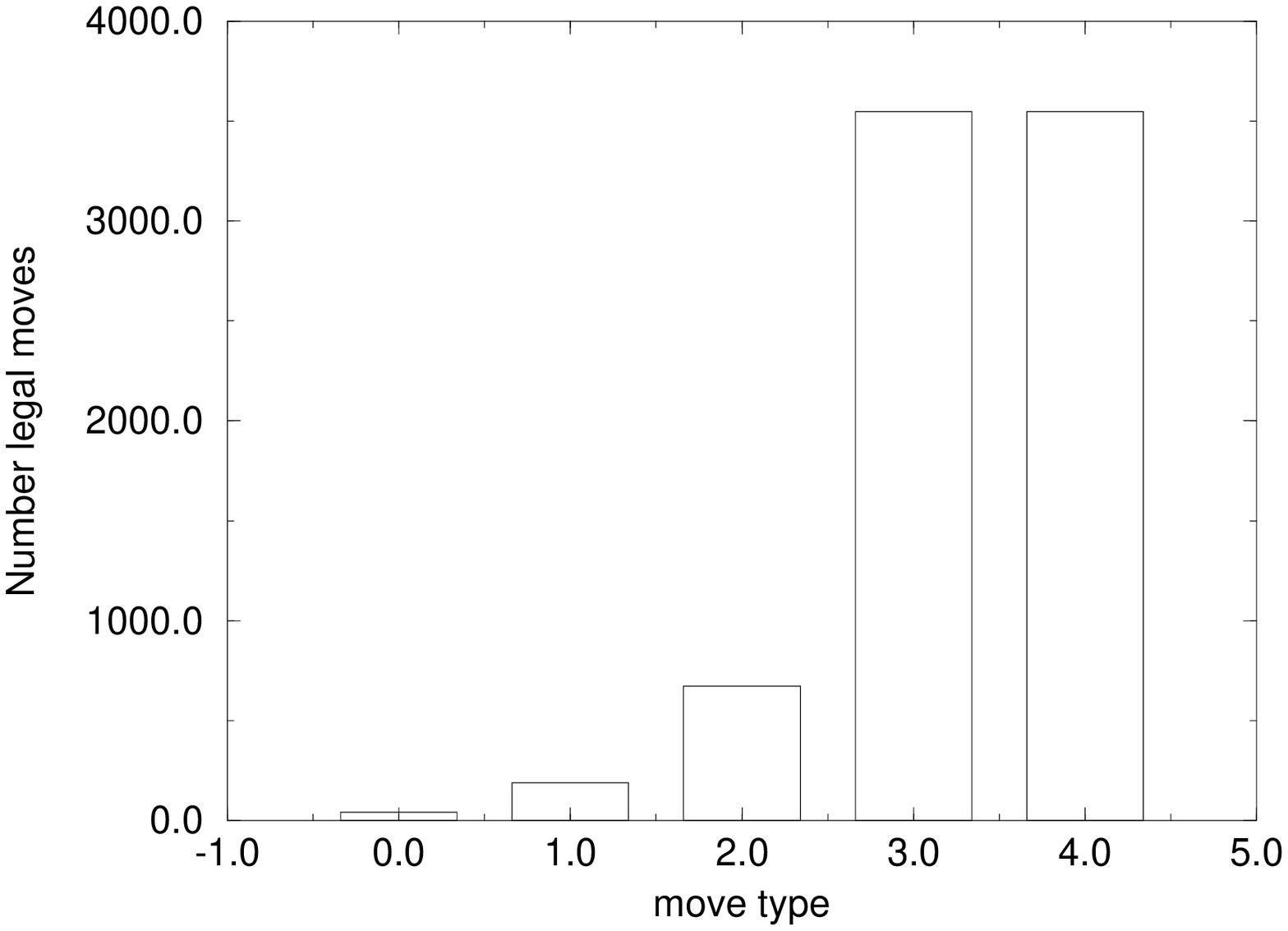}
\caption{Number legal subsimplices per sweep $d=4$, $V=8000$, $\kappa_0=0$}
\label{fig2}
\end{center}
\end{figure}

\begin{figure}
\begin{center}
\leavevmode
\epsfxsize=400pt
\epsfbox{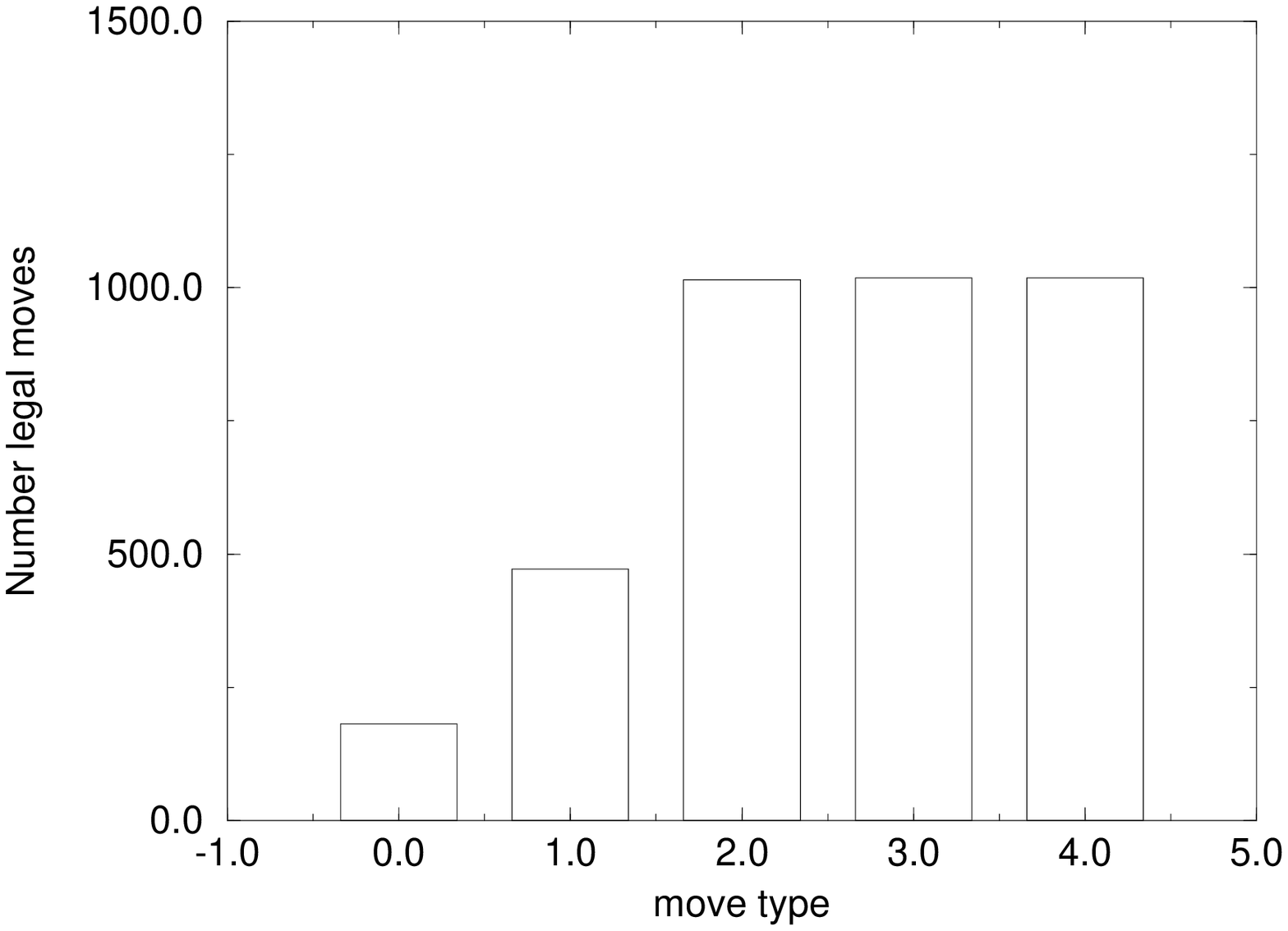}
\caption{Number legal subsimplices per sweep $d=4$, $V=8000$, $\kappa_0=2.4$}
\label{fig3}
\end{center}
\end{figure}

For the same run, fig. \ \ref{fig2} shows $L\left(i\right)$ 
the average number of 
legal subsimplices
of type $i$ encountered per sweep. Clearly, by the definition of
the triangulated manifold, $3$ and $4$ subsimplices are always legal,
whilst at this value of $\kappa_0=0$ there are relatively few
legal nodes and links (i.e 5-fold coordinated vertices and 4-fold
coordinated links).  This is to be contrasted with fig. \ \ref{fig3}
where the same quantity is plotted for $\kappa_0=2.4$. The number
of legal nodes has increased by nearly a factor of five.

\begin{figure}
\begin{center}
\leavevmode
\epsfxsize=400pt
\epsfbox{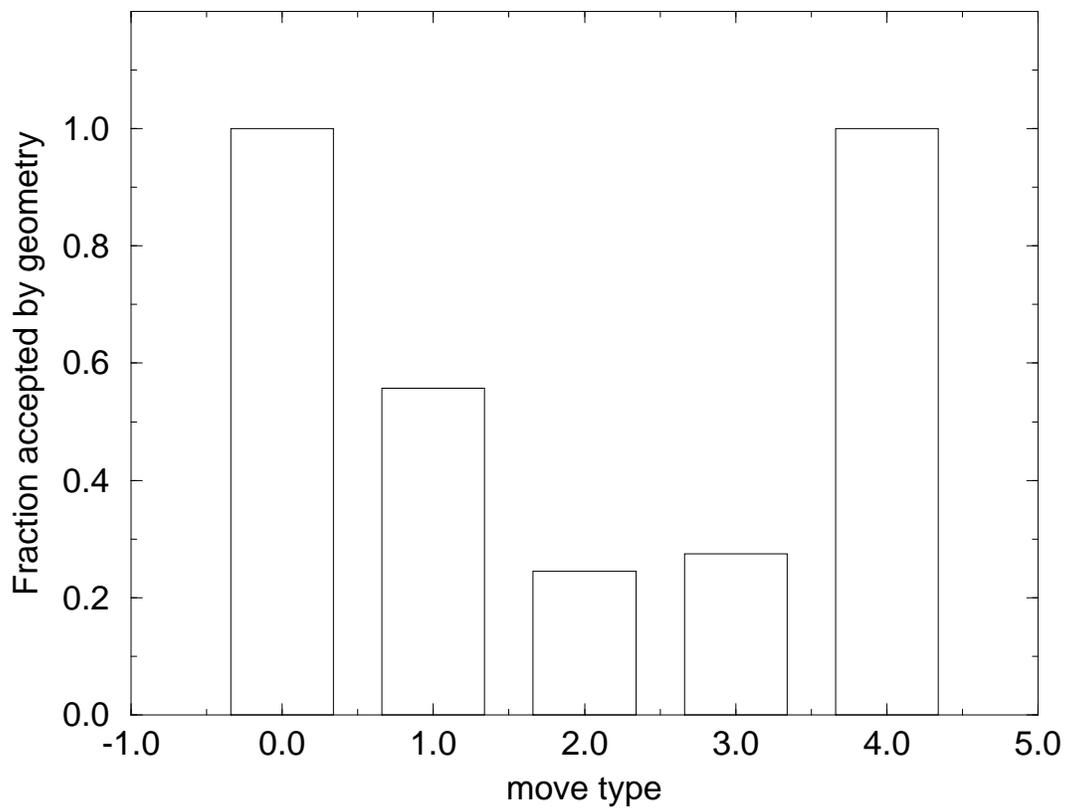}
\caption{Fraction legal i-moves allowed geometrically, $d=4$, $V=8000$}
\label{fig4}
\end{center}
\end{figure}

In fig. \ \ref{fig4} we show the fraction of legal subsimplices 
$P\left(i\right)$ for which
an update would lead to an geometrically acceptable triangulation.
Again the lattices are four dimensional with mean volume $V=8000$ at
zero node coupling.
In the case of move $0$ (node deletion) this
is possible with unit probability $P\left(0\right)=1$, but for
subsequent moves $P\left(i\right)$ decreases until for moves of type $4$ (node
insertion) it again reaches unity. 

\begin{figure}
\begin{center}
\leavevmode
\epsfxsize=400pt
\epsfbox{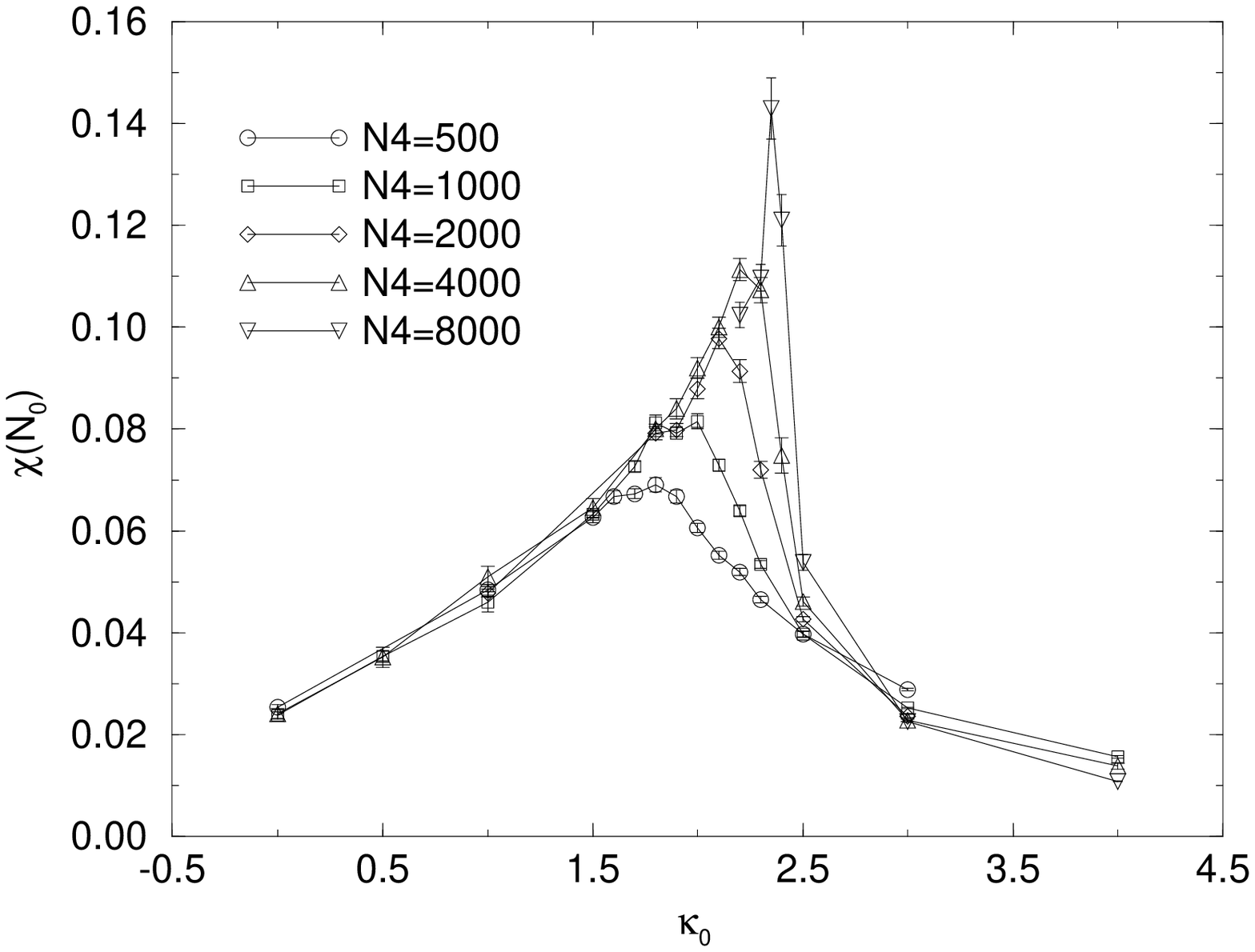}
\caption{Node susceptibility $d=4$}
\label{fig5}
\end{center}
\end{figure}

As evidence of the two phase structure mentioned in the introduction
we show in fig. \ \ref{fig5} a plot of the node susceptibility $\chi$ as
a function of node coupling $\kappa_0$ and
for a variety of four dimensional lattice volumes $V=500-8000$. 

\begin{equation}
\chi={1\over V}\left(\left\langle N_0^2\right\rangle-\left\langle N_0\right
\rangle^2\right)
\end{equation}

There appears to be
a growing peak which shifts and narrows with increasing volume. For
conventional statistical mechanical systems this would be taken as
an indicator of a phase transition. This quantity is
sensitive to the presence of long range correlations in
the geometric curvature.
The node coupling being (inversely) related to a bare Newton 
gravitational constant, this is taken as evidence that there may be
a nontrivial fixed point in the theory about which it may be
possible to have a consistent quantum theory of euclidean gravity.

\section*{Conclusions}

We have described, in some detail, an algorithm to simulate models
for quantum gravity based on dynamical triangulations. We have demonstrated
that the necessary procedures can be implemented in such a way that
their dependence on dimension is trivial - a single compact code can
be written in which the dimension is simply input as a parameter.

Furthermore, we have illustrated the utility of such a code by
recording the results of some high statistics studies of the four
dimensional theory. Numerical evidence is presented to support
a phase transition. 

SMC would like to acknowledge useful conversations with John Kogut
and Ray Renken.

\vfill
\newpage

\vfill
\newpage
\begin{thebibliography}{99}

\bibitem{mig} D. Boulatov, V. Kazakov, I. Kostov and A. Migdal, Nucl. Phys.
B275 (1986) 641.
\bibitem{amb} J. Ambj\o rn, B. Durhuus and J. Frohlich, Nucl. Phys. B257 
(1985) 433.
\bibitem{dav} F. David, Nucl. Phys. B257 (1985) 543.
\bibitem{kpz} V. Knizhnik, A. Polyakov and A. Zamolodchikov, Mod. Phys. Lett.
A3 (1988) 819.
\bibitem{ddk} J. Distler and H. Kawai, Nucl. Phys. B321 (1989) 509.
\bibitem{mig2d} M. Agishtein and A. Migdal, Nucl. Phys. B350 (1991) 690.
\bibitem{jurkis} J. Jurkiewicz, A. Krzywicki, B. Petersson and B. Soderberg,
Phys. Lett. B213 (1988) 511.
\bibitem{usis} S. Catterall, J. Kogut and R. Renken, Phys. Rev. D45 (1992) 2957.
\bibitem{clivepot} C. Baillie and D. Johnston, Mod. Phys. Lett. A7 (1992) 1519.
\bibitem{kaz} V. Kazakov, Phys. Lett. B119 (1986) 140.
\bibitem{puregrav} V. Kazakov and A. Migdal, Nucl. Phys. B311 (1989) 171.
\bibitem{multis} S. Catterall, J. Kogut and R. Renken, Phys. Lett. B292 (1992)
277.
\bibitem{multipotts} C. Baillie and D. Johnston, Phys. Lett. B286 (1992) 44.
\bibitem{multiamb} G. Thorleifsson, Lattice 92, Nucl. Phys. B (Proc. Suppl.) in
press.
\bibitem{xy} S. Catterall, R. Renken and J. Kogut, ILL-(TH)-93-5, Nucl. Phys. B
in press.
\bibitem{3dgross} N. Godfrey and M. Gross, Phys. Rev. D43 (1991) R1749.
\bibitem{3damb} J. Ambj\o rn and S. Varsted, Nucl. Phys. B373 (1992) 557.
\bibitem{3dmig} M. Agistein and A. Migdal, Mod. Phys. Lett. A6 (1991) 1863.
\bibitem{3dboul} D. Boulatov and A. Krzywicki, Mod. Phys. Lett. A6 (1991) 3005.
\bibitem{4damb} J. Ambj\o rn and J. Jurkiewicz, Phys. Lett. B278 (1992) 42.
\bibitem{4dsteen} S. Varsted, UCSD/PTH 92/03.
\bibitem{4dmig} M. Agistein and A. Migdal, Mod. Phys. Lett. A7 (1992) 1039.,
Nucl. Phys. B385 (1992) 395.
\bibitem{4dbrug} B. Brugmann and E. Marinari, Phys. Rev. Lett. 70 (1993) 1908.
\bibitem{4dus} S. Catterall, J. Kogut and R. Renken, CERN-TH.7149/94, 
Phys. Lett. B in press.
\bibitem{gross} M. Gross and S. Varsted, Nucl. Phys. B378 (1992) 367.
\bibitem{ergod} A. Nabutovsky and R. Ben-Av, PUPT-1327.
\bibitem{bound} S. Catterall, J. Kogut and R. Renken, CERN-TH.7197/94,
Phys. Rev. Lett in press.

\end{thebibliography}
\end{document}